\documentclass[prb,superscriptaddress, floatfix, letterpaper,twocolumn,aps,showpacs]{revtex4-2}
\usepackage[utf8]{inputenc}
\usepackage{natbib}
\usepackage{amsmath}
\usepackage{amsbsy}
\usepackage{amssymb}
\usepackage{scrextend}
\usepackage{graphicx}
\usepackage{color}
\usepackage{xcolor}
\usepackage{float}
\usepackage[colorlinks=true, allcolors = blue]{hyperref}

\usepackage{array} 
\newcolumntype{L}[1]{>{\raggedright\let\newline\\\arraybackslash\hspace{0pt}}m{#1}}
\newcolumntype{C}[1]{>{\centering\let\newline\\\arraybackslash\hspace{0pt}}m{#1}}
\newcolumntype{R}[1]{>{\raggedleft\let\newline\\\arraybackslash\hspace{0pt}}m{#1}}

\renewcommand{\vec}[1]{\boldsymbol{#1}}
\newcommand{\BaKx}{{$\mathrm{Ba}_{1-x}\mathrm{K}_{x}\mathrm{Fe}_{2}\mathrm{As}_{2}$}}

\begin{document}
\title{Interplay of stripe and double-Q magnetism with superconductivity in {\BaKx} under the influence of magnetic fields}
\author{K.\ Willa}
\email{kristin.willa@kit.edu}
\affiliation{Institute for Quantum Materials and Technologies, Karlsruhe Institute of Technology, Karlsruhe D-76021, Germany}

\author{R.\ Willa}
\affiliation{Institute for Condensed Matter Theory, Karlsruhe Institute of Technology, Karlsruhe D-76131, Germany}
\author{F.\ Hardy}
\affiliation{Institute for Quantum Materials and Technologies, Karlsruhe Institute of Technology, Karlsruhe D-76021, Germany}
\author{L.\ Wang}
\affiliation{Institute for Quantum Materials and Technologies, Karlsruhe Institute of Technology, Karlsruhe D-76021, Germany}
\author{P.\ Schweiss}
\affiliation{Institute for Quantum Materials and Technologies, Karlsruhe Institute of Technology, Karlsruhe D-76021, Germany}
\author{T.\ Wolf}
\affiliation{Institute for Quantum Materials and Technologies, Karlsruhe Institute of Technology, Karlsruhe D-76021, Germany}
\author{C.\ Meingast}
\affiliation{Institute for Quantum Materials and Technologies, Karlsruhe Institute of Technology, Karlsruhe D-76021, Germany}

\date{\today}

\begin{abstract}
At $x\approx0.25$ {\BaKx} undergoes a novel first-order transition from a four-fold symmetric double-Q magnetic phase to a two-fold symmetric single-Q phase, which was argued to occur simultaneously with the onset of superconductivity (Böhmer et al., Nat. Comm. 6, 7911 (2015)). Here, by applying magnetic fields up to 10T, we investigate in more detail the interplay of superconductivity with this magneto-structural transition using a combination of high-resolution thermal-expansion and heat-capacity measurements. We find that a magnetic field suppresses the reentrance of the single-Q orthorhombic phase more strongly than the superconducting transition, resulting in a splitting of the zero-field first-order transition. The suppression rate of the orthorhombic reentrance transition is stronger for out-of-plane than for in-plane fields and scales with the anisotropy of the superconducting state. These effects are captured within a phenomenological Ginzburg-Landau model, strongly suggesting that the suppression of the reentrant orthorhombic single-Q phase is primarily linked to the field-induced weakening of the superconducting order. Not captured by this model is however a strong reduction of the orthorhombic distortion for out-of-plane fields, which deserves further theoretical attention.
\end{abstract}

\maketitle
\section{Introduction}
\begin{figure*}[tbh]
\includegraphics[width=.95\linewidth]{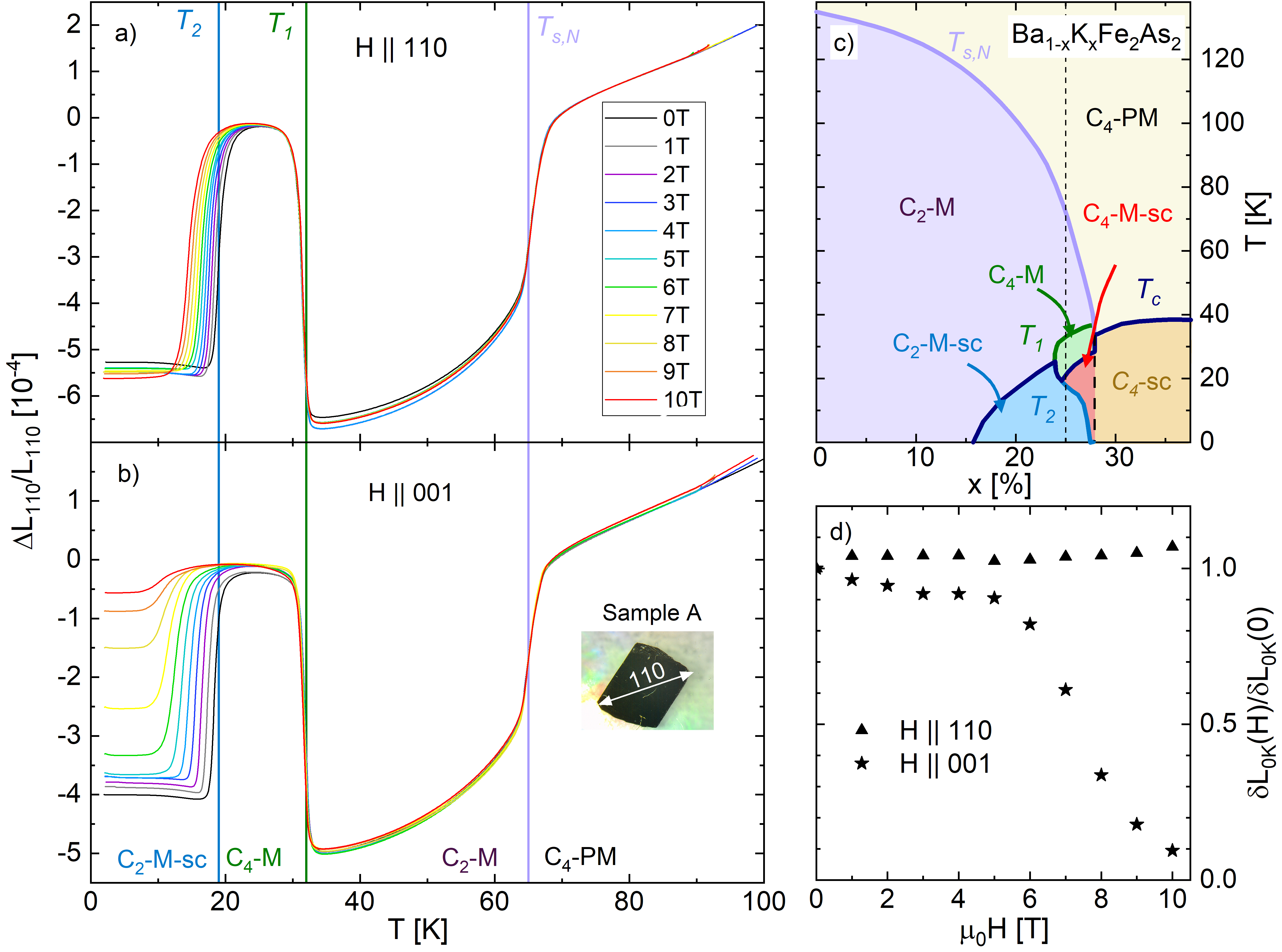}
\caption{Temperature dependence of the thermal expansion $\Delta L / L$ of {\BaKx} measured along the [110] direction in different external magnetic fields where for a) H $\parallel$ [110] and for b) H $\parallel$ [001]. The discontinuities in $\Delta L/L$ mark the transitions between the $C_2$ and $C_4$-phases as indicated at the bottom of the figure. c) Phase diagram of {\BaKx} for $0<x<0.3$ \cite{Bohmer2015}, showing the narrow doping region in which a C4-phase is observed. d) The extrapolated magnetic field dependent step height $\delta L_{0K}(H)/\delta L_{0K}(0)$ as extracted from the thermal expansion measurements. 
}
\label{fig:ThermalExpansion}
\end{figure*}

Coexistence of superconductivity and magnetism as well as their competition for phase space is a recurring characteristic of iron-based superconductors \cite{Lorenzana2008, Fernandes2014}. Superconductivity emerges when the prevailing stripe-type antiferromagnetic spin density wave (SDW)---is suppressed by either hole/electron doping or pressure \cite{Paglione2010, Rotter2008a, Torikachvili2008}. This SDW state is accompanied by an orthorhombic distortion of the lattice which occurs simultaneously or sometimes precedes it in form of a vestigial nematic transition. Moreover, careful investigations of hole-doped systems, including $\mathrm{Ba}_{1-x}\mathrm{Na}_{x}\mathrm{Fe}_{2}\mathrm{As}_{2}$, $\mathrm{Ba}_{1-x}\mathrm{Ka}_{x}\mathrm{Fe}_{2}\mathrm{As}_{2}$, $\mathrm{Sr}_{1-x}\mathrm{Na}_{x}\mathrm{Fe}_{2}\mathrm{As}_{2}$ and $\mathrm{Ca}_{1-x}\mathrm{Na}_{x}\mathrm{Fe}_{2}\mathrm{As}_{2}$, have revealed a plethora of competing new electronic phases near the transition region between stripe antiferromagnetism ($C_2$) and superconductivity. In particular a tetragonal ($C_4$) magnetic phase was discovered in a narrow doping region \cite{Avci2014, Bohmer2015, Wang2016, Taddei2017, Wang2019}. Neutron studies \cite{Allred2015, Wasser2015} showed that in this magnetically ordered $C_4$ phase, the moments flip from in-plane to out-of-plane and M\"ossbauer studies as well as Muon spin rotation measurements \cite{Mallett2015, Allred2016} revealed a double-$Q$ order, in which only every other Fe atom carries a magnetic moment. According to the classification of possible double-$Q$ magnetic orders (including, hedgehog and loop spin-vortex crystals), see Ref.~[\onlinecite{Meier2018}], this finding is consistent with the so-called spin-charge density wave.
In the case of {\BaKx}, initial studies \cite{Hassinger2012, Avci2011, Avci2012, Tanatar2014} did not find the small doping region where the double-Q $C_4$-phase occurs, which was only later revealed upon closer investigation \cite{Bohmer2015}, see Figure \ref{fig:ThermalExpansion}~c). In contrast to $\mathrm{Ba}_{1-x}\mathrm{Na}_{x}\mathrm{Fe}_{2}\mathrm{As}_{2}$ and $\mathrm{Sr}_{1-x}\mathrm{Na}_{x}\mathrm{Fe}_{2}\mathrm{As}_{2}$ \cite{Wang2016, Wang2019, Sheveleva2020}, the $C_4$ phase in {\BaKx} does not extent to zero temperature; rather superconductivity surprisingly drives the system back to a single-Q orthorhombic magnetic phase \cite{Bohmer2015, Timmons2019}. Böhmer et al. proposed that this was due to the higher electronic entropy available for superconductivity in the single-Q compared to the double-Q phase. 


Here, by applying magnetic fields up to 10 T, we investigate in more detail the interplay of superconductivity with this magnetostructural transition using a combination of high-resolution thermal-expansion and heat-capacity measurements. We find that a magnetic field suppresses the reentrance of the single-Q orthorhombic phase more strongly than the superconducting transition, resulting in a splitting of the zero-field first-order transition. The suppression rate of the orthorhombic reentrance transition, which remains first-order, is stronger for out-of-plane than for in-plane fields and scales with the anisotropy of the superconducting state.  These effects are captured within a phenomenological Ginzburg-Landau model, strongly suggesting that the suppression of the reentrant orthorhombic magnetic phase is primarily linked to the field-induced weakening of the superconducting order.  Not captured by this model is however the strong reduction of the orthorhombic distortion for out-of-plane fields, which deserves further theoretical attention.


\begin{figure*}[tbh]
\includegraphics[width=0.95\linewidth]{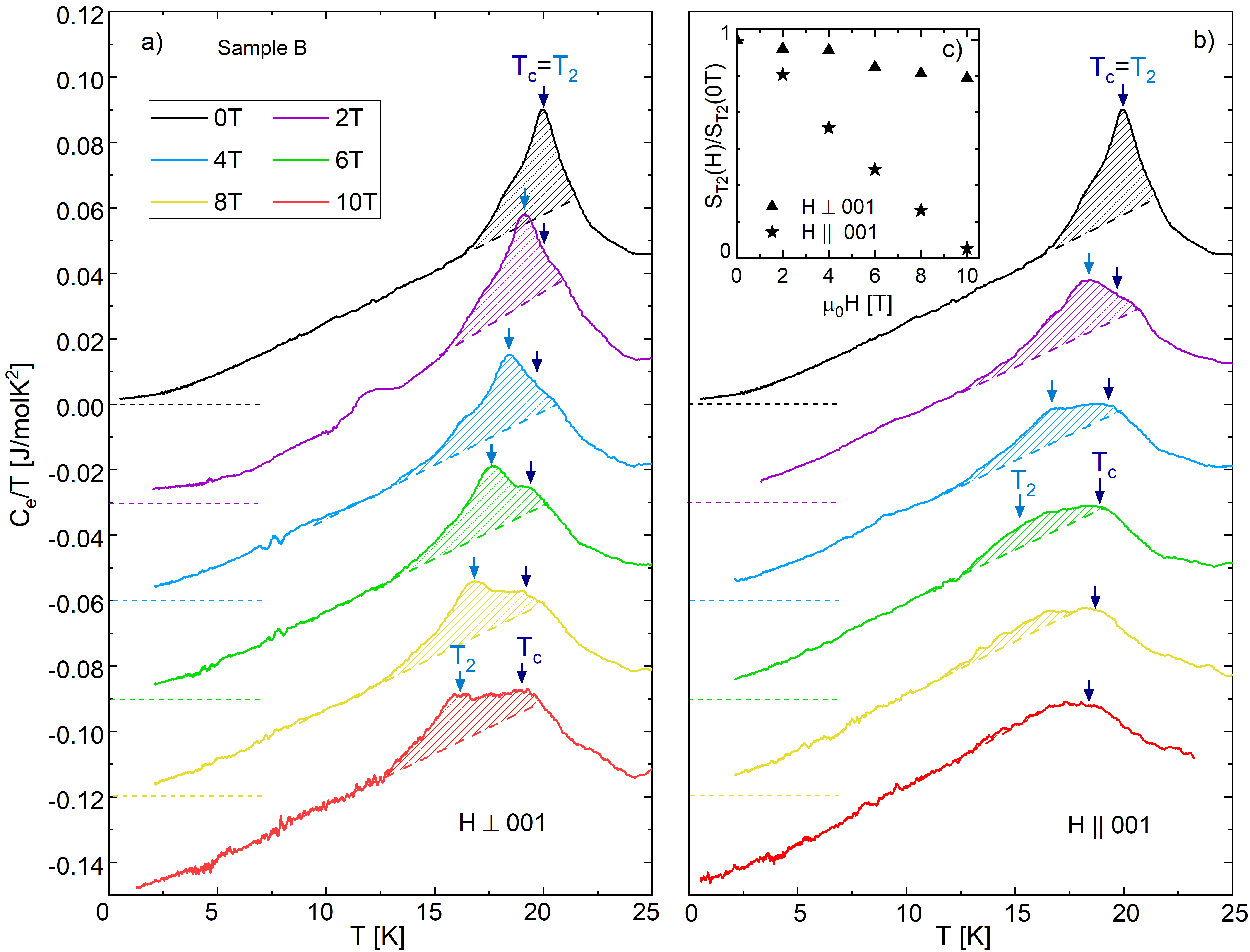}

\caption{Temperature dependence of the electronic specific heat for {\BaKx} subject to different applied magnetic fields up to 10T. The left (right) panel shows the response to in-plane (out-of-plane) magnetic fields. The arrows mark the location of the superconducting and structural transitions, respectively. The shaded region highlights the entropy corresponding to the structural distortion. Curves are shifted vertically by -0.03J/molK$^2$ for every 2T for clarity, the dashed lines mark the corresponding $C/T=0$. The inset c) shows the entropy associated to the $C_4$-to-$C_2$ phase transition (equal to the shaded region in $C/T$) normalized by the zero-field entropy.
}
\label{fig:spec_heat}
\end{figure*}

\section{Experimental details}

Heat-capacity and thermal expansion measurements were made on two platelet shaped self-flux grown single crystals $\mathrm{Ba}_{1-x}\mathrm{K}_{x}\mathrm{Fe}_{2}\mathrm{As}_{2}$ ($x \approx 0.24$) with masses of 2.63mg (Sample A) and 1.93mg (Sample B), in plane dimensions of $2.1\!\times\!2.6 \mathrm{mm}^{2}$ (Sample A). The samples were grown by a self-flux technique \cite{Bohmer2015}. Both crystals are from the same batch as in Ref. \onlinecite{Bohmer2015} and where more details about the crystal synthesis are provided. Thermal-expansion measurements were performed on Sample A in a home-made high-resolution capacitive dilatometer \cite{Meingast1990}. As demonstrated previously, measurements of the crystal along the [110]$_T$ direction of the original tetragonal cell ($C_4$-PM) produces partial 'detwinning' due to the pressure of the springs in the dilatometer cell, giving access to the thermal expansion of the shorter b-axis in the low-temperature stripe phase (C$_2$-M) \cite{Bohmer2015, Wang2016, He2018}. Rotating the dilatometry cell allows to align the magnetic field along both the [110]$_T$ and the [001] crystal directions. Specific heat measurements were performed on sample B in a 14T Physical Property Measurement system from Quantum Design with the heat-capacity option. 

\section{Results}

Figures \ref{fig:ThermalExpansion}a) and \ref{fig:ThermalExpansion}b) show the relative thermal expansion $\Delta L_{110}/L_{110}$ of sample A for fields applied parallel to the [110]$_T$ and [001] directions, respectively. In zero field, a pronounced reduction of $\Delta L_{110}/L_{110}$ occurs at $T_{s,N} = 65$K indicative of the transition from the paramagnetic tetragonal state into the $C_{2}$-magnetic stripe phase. At lower temperatures, a switch back of $\Delta L_{110}/L_{110}$ is observed at $T_1=32$K, where the system enters the double-Q magnetic phase ($C_4$-M) that restores tetragonal symmetry, followed by another sudden reduction at $T_2=19$K marking the simultaneous re-entrance of stripe magnetism and the emergence of superconductivity. Application of an external magnetic field of up to 10T does not affect the magnetic transitions at $T_{s,N}$ and $T_1$ for either field direction, indicative of a very robust magnetism. The locus of these first-order phase transitions and the determined $\mathrm{K}$ concentration of sample A are in good agreement with the reported $(x,T)$ phase diagram of Ref. \onlinecite{Bohmer2015}, schematically reproduced in Fig. \ref{fig:ThermalExpansion}c) (dashed line). The reduced value of $\Delta L_{110}/L_{110}$ for $T<T_2$ with respect to that from extrapolation from the $C_2$-M stripe phase indicates that superconductivity, which favors the single-Q stripe phase over the double-Q phase, however still competes with the former \cite{Nandi2010}.
In contrast, to the transitions at $T_{s,N}$ and $T_1$, a clear shift of $T_{2}(H)$ towards lower temperature is observed with increasing field for both field orientations. However, the opposite evolution of $\Delta L_{110}/L_{110}$, which is an indicator of the orthorhombic distortion, is quite different inside this re-entrant $C_2$ phase for both field directions. Whereas $\Delta L_{110}/L_{110}(H)$ is slightly enhanced for fields along [110]$_T$, it is drastically reduced to $\approx 10\%$ of its zero-field value for $\mu_0H=10$T along the [001] direction, as illustrated in figure \ref{fig:ThermalExpansion}d), suggesting an almost complete suppression of the reentrant $C_2$ phase. These opposite behaviors with field orientation cannot be solely accounted for by the re-entrance of the stripe phase, and the influence of superconductivity, which emerges concomitantly in zero field, must be considered. In order to gain further insight about the interplay of superconductivity with the reentrant transition, we have also measured the heat capacity for both field orientations. 

Figures \ref{fig:spec_heat}a) and \ref{fig:spec_heat}b) show the respective electronic specific heat of sample B for fields applied perpendicular and parallel to the [001] direction, obtained after subtracting a suitable lattice contribution (see e.g. Refs.~\cite{Bohmer2015, Hardy2014, Hardy2010, Hardy2016} for details). In zero field, a single first-order-like peak at $T_c=T_2 = 21$K confirms that the re-entrance into the $C_2$ stripe phase occurs simultaneously with superconductivity, as observed in Ref.\ \onlinecite{Bohmer2015}. This is corroborated by the vanishing $C_e/T$ in the $T\rightarrow 0$ limit, indicative of a fully-gapped Fermi surface.

With increasing field, a progressive splitting of the zero-field peak into two distinct anomalies is unambiguously resolved for $H\perp$[001]. Here, a broadened mean-field-like anomaly at $T_c(H)$ marks the transition to the superconducting state (dark blue arrow) followed by a peak at $T_2(H)<T_c(H)$ where reentrance of the stripe phase takes place (light blue arrow). Thus, our data reveal that the application of a magnetic field causes the apparition of two distinct superconducting states coexisting with either the single-Q stripe SDW or the double-Q C-SDW, separated by the $T_2(H)$ boundary. In field, the transition at $T_2$ broadens somewhat, but remains first-order as illustrated by the entropy discontinuity still observed in 10 T (see Fig. \ref{fig:phaseboundary}a)). This is also consistent with the discontinuity observed in the thermal expansion at $T_2$. For $H\parallel$[001], the splitting is also clearly observed (see Fig.\ref{fig:spec_heat}b)), but with a more pronounced suppression of $T_2(H)$. In contrast to in-plane fields, the peak at $T_2 (H)$ (blue arrow) is rapidly suppressed with increasing $c$-axis fields and fades away for $H>6$T. We have estimated the entropy associated with this transition by the area between the measured curve and an extrapolation of the low-T quasi linear behavior to $T=T_c$, see shaded areas in Figure~\ref{fig:spec_heat}. Most prominently, we note that this entropy discontinuity rapidly decreases with increasing field strength for fields along the $c$ axis, while experiencing only moderate changes for in-plane fields. This trend, shown in the inset of Figure~\ref{fig:spec_heat}, is in line with the field dependence of the orthorhombic distortion, as inferred from our thermal-expansion data, Fig.~\ref{fig:ThermalExpansion}d). We note that we do not observe a clear signature at $T_c$ in the thermal expansion even at 10 T, where there is a clear separation between $T_2$ and $T_c$. This is surprising, since, although small, clear thermal expansion anomalies at $T_c$ have been observed in both $\mathrm{Ba}_{1-x}\mathrm{Na}_{x}\mathrm{Fe}_{2}\mathrm{As}_{2}$ and $\mathrm{Sr}_{1-x}\mathrm{Na}_{x}\mathrm{Fe}_{2}\mathrm{As}_{2}$ in the double-Q phase \cite{Wang2016, Wang2019}.

\begin{figure*}[htb]
\includegraphics[width=0.9\linewidth]{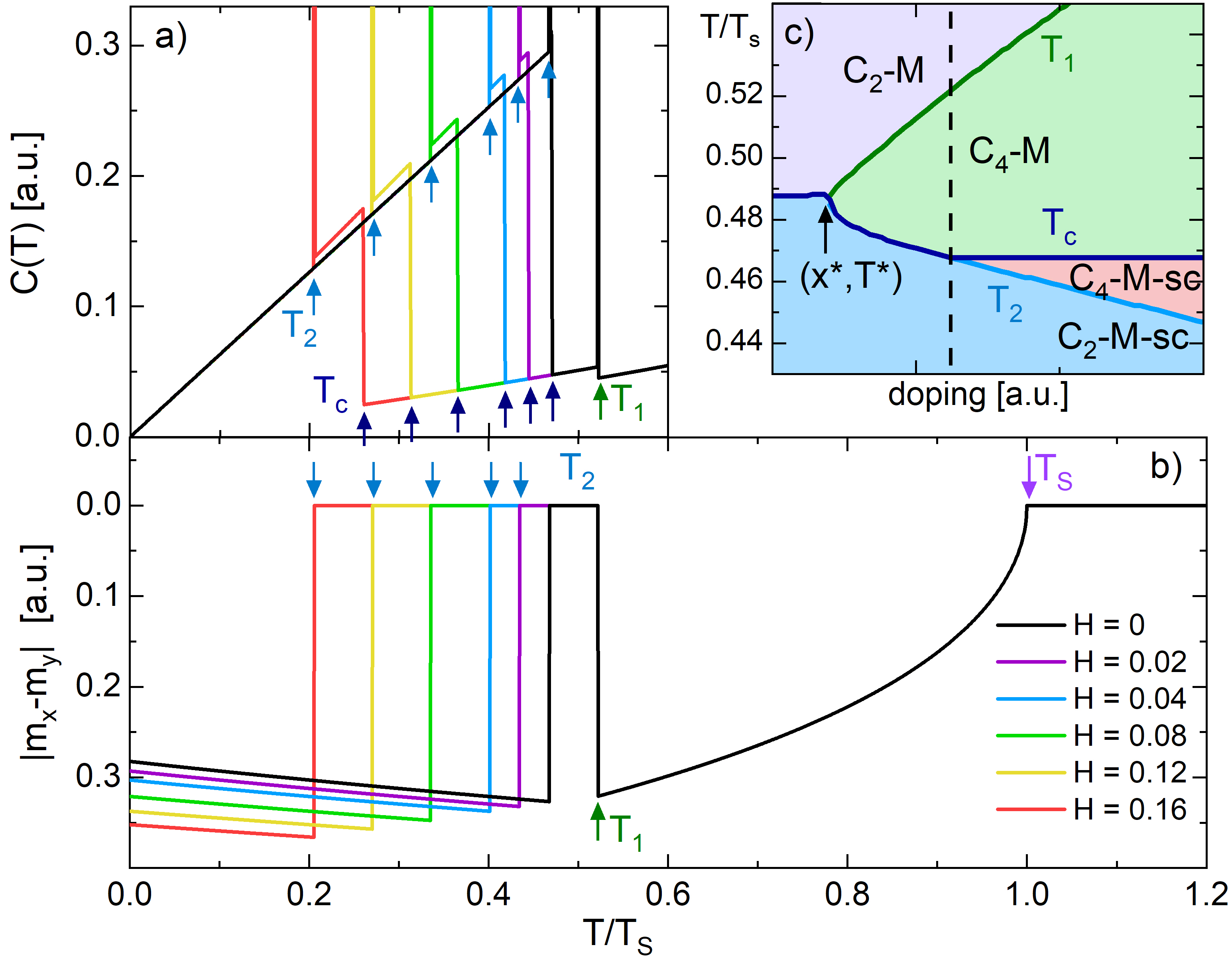}
\caption{a) Calculated heat capacity near the onset of superconductivity for different magnetic field strengths. At zero field, the transition is first-order and accompanied by a magnetic switching from the $C_{4}$ to the $C_{2}$ stripe order. For sufficiently large fields, $H \gtrsim 0.02$, this transition is split into a second order onset of superconductivity, followed by a first-order magnetic switching. The proxy $|m_{x} - m_{y}|(T)$ of the orthorhombic distortion for different magnetic fields is shown below in b). Transition temperatures are marked with arrows. c) Close-up of the phase diagram near the tricritical point $(x^{*},T^{*})$ where the $C_{2}$ and $C_{4}$ magnetic phases are degenerate and where superconductivity appears. For larger doping the first-order onset of superconductivity is accompanied by a transition into a $C_{2}$ magnetic phase. At even higher doping levels, these two transitions split and allow for a superconducting phase with $C_{4}$ magnetic order. The dashed line marks the doping concentration for which the specific heat and orthorhombic distortion are shown in a) and b).
}
\label{fig:simulations}
\end{figure*}

\begin{figure*}[t!]
\includegraphics[width=0.98\textwidth]{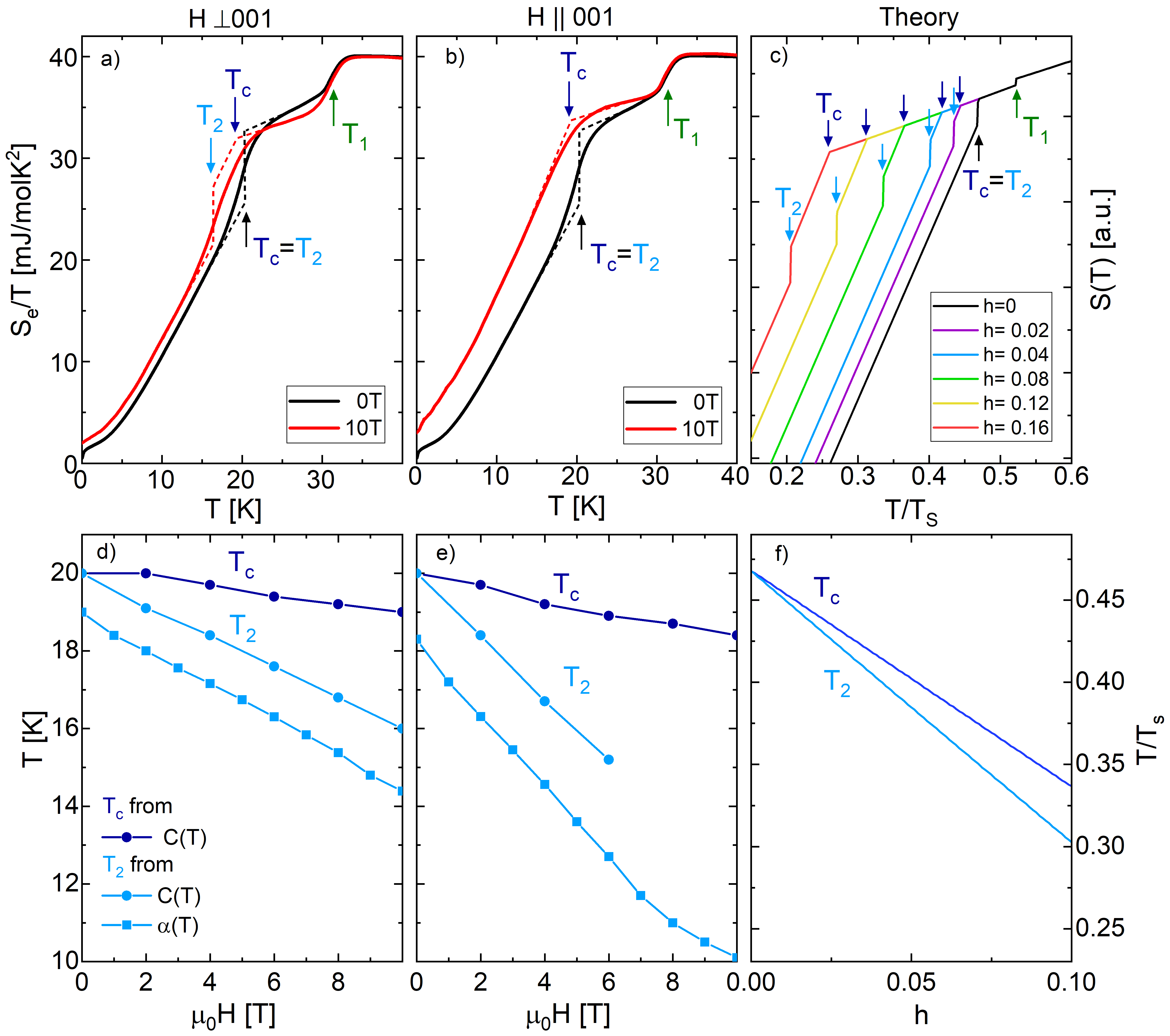}
\caption{Panels a) and b) show the temperature dependence of the electronic entropy in 0T and for 10T applied in- and out-of-plane fields. Panel c) shows the calculated entropy for the theoretical model. $T_c$, $T_2$, and $T_1$ are marked with arrows. The dashed lines guide the eye along the theoretical entropy curve. Panels d) and e) show the low temperature phase diagram with transition temperatures $T_c$ and $T_2$ taken from specific heat (circles) and thermal expansion (rectangles). f) Theoretically calculated phase boundary.
}
\label{fig:phaseboundary}
\end{figure*}

\section{Theoretical model}
Multiple theoretical studies \cite{Christensen2015, Christensen2018, Christensen2017, Scherer2018} have investigated the interplay between the different possible magnetic ground states in iron-based superconductors.
For a qualitative understanding of the physical mechanisms underlying the presently observed magnetic field dependence of the reentrant transition in {\BaKx}, we aim at capturing the key observations within a minimal phenomenological model. We start from the treatment proposed by Kang \emph{et al.} \cite{Kang2015}, and account for a spin-density modulation $m = m_{x} e^{i \vec{Q}_{x} \vec{r}} + m_{y} e^{i \vec{Q}_{y} \vec{r}}$ and the leading $s_{\pm}$ superconducting pairing state $\Delta$ \cite{Kuroki2009, Maiti2011, Bohm2014}. 
The effective theory for the SDW fields $\vec{m}_{x/y} \!\propto\! \frac{1}{2}\sum_{\vec{k}} c^{\dagger}_{\vec{k}} \vec{\sigma} c_{\vec{k}+\vec{Q}_{x/y}}$ ($m_{x/y} \!=\! |\vec{m}_{x/y}|$) is obtained by applying a Hubbard-Stratonovich transformation to decouple the fermion interaction and subsequently integrating out the fermionic fields. The free energy functional
$\mathcal{F}[m_{x},m_{y},\Delta] = \mathcal{F}_{m} + \mathcal{F}_{s} + \mathcal{F}_{\mathrm{int}}$ consists of a magnetic part
\begin{align}\label{eq:Fm}
   \!\!\!
   \mathcal{F}_{m} &= \frac{\alpha}{2} (m_{x}^{2} + m_{y}^{2})
                     + \frac{u}{4} (m_{x}^{2} + m_{y}^{2})^{2}
                     - \frac{g}{4} (m_{x}^{2} - m_{y}^{2})^{2} \nonumber\\
                     &\quad
                     + \frac{v}{6} (m_{x}^{2} + m_{y}^{2}) (m_{x}^{2} - m_{y}^{2})^{2}
                     + \frac{\gamma}{6} (m_{x}^{2} + m_{y}^{2})^{3},
\end{align}
a minimal superconducting contribution
\begin{align}\label{eq:Fs}
   \mathcal{F}_{s} &= \frac{\alpha_{s}}{2} \Delta^{2} + \frac{u_{s}}{4} \Delta^{4},
\end{align}
and a mutual interaction
\begin{align}\label{eq:Fint}
   \mathcal{F}_{\mathrm{int}} &= \frac{c_{1}}{2} \Delta^{2}(m_{x}^{2} + m_{y}^{2})
                                 + 2 c_{2} \Delta^{2} m_{x}^{2} m_{y}^{2}.
\end{align}
The magnetic sector $\mathcal{F}_{m}$ allows either for a $C_{4}$ order with $m_{x} = m_{y}$, or two degenerate $C_{2}$ orders with $m_{x}=m$ and $m_{y}=0$ or vice versa. The energy sheets associated with these two possibilities evolve differently upon changing external parameters (temperature, doping) and will trigger first-order transitions whenever they cross. Upon entering the magnetic phase, the term $- (g/4) (m_{x}^{2} - m_{y}^{2})^{2}$ favors the $C_{2}$ magnetic order for $g > 0$. For $v > 0$ the sixth-order term may tip the balance in favor of the $C_{4}$ phase. Consequently, $v$ seems a suitable tuning parameter to capture the dominant doping dependence. The first interaction term in $\mathcal{F}_{\mathrm{int}}$ accounts for a competition between superconductivity and magnetism ($c_1>0$). The second term ($c_2>0$) further enhances the competition of superconductivity with the $C_4$-magnetic order while being absent within the $C_2$-magnetic phase.
Our model extends the one of Kang \emph{et al.} \cite{Kang2015} by including a magnetic field dependence in $\alpha_{s}$. Owing to the magnetic field robustness of the first two magnetic transitions observed in the experiment, we attribute the leading field-dependence to the superconducting orbital pair-breaking. In addition, the second term in $\mathcal{F}_{\mathrm{int}}$ brings a strong energetic distinction on how the two magnetic orders couple to superconductivity. Such an interaction has been shown to arise from the coupling of the superconducting $s_{\pm}$ order to an underlying, yet unfulfilled, $d$-wave gap function \cite{Kang2015}. When the system realizes the $C_{4}$ phase with a nearly degenerate (but subordinate) $C_{2}$ state, the onset of superconductivity can trigger a first-order transition into a superconducting $C_{2}$ state. More details on the model and the specific parameter values are given in Appendix~\ref{app:model}.

The phase diagram in zero magnetic field, Figure~\ref{fig:simulations}~c), reproduces well the observation reported in Ref.~\cite{Bohmer2015}. Upon cooling, the system first enters the $C_{2}$ stripe phase, followed, at low doping, by the onset of superconductivity. Above a critical doping, the $C_{2}$-magnetic state undergoes a first-order transition into a tetragonal double-$Q$ magnetic phase which lowers the onset of superconductivity, presumably due to additional gapping of the fermi surface in the double-Q phase \cite{Bohmer2015}. This lost density of states is recovered when, with the onset of superconductivity, the magnetism switches back to the $C_2$ stripe phase. At even larger doping levels, the last two transitions appear sequentially, with superconductivity setting in as a second-order transition, followed by a first-order switching to the $C_{2}$-magnetic state.

For a fixed doping level, indicated by the dashed line in figure \ref{fig:simulations}c), we have calculated the heat capacity and $|m_{x} - m_{y}|(T)$ as a proxy for the orthorhombic lattice distortion. Both quantities associated to our model calculation are shown in Figure~\ref{fig:simulations}a) and b). In zero magnetic field, we observe the first order transition into the double-Q charge density wave at $T_1$ and then superconductivity drives the system back in to a single-Q SDW. This joint transition is well captured by both a delta peak and a step in the heat capacity as well as a step in the lattice distortion. In magnetic fields, the transition splits apart and the system becomes first superconducting (step in $C(T)$) and then orthorhombic (delta peak in $C(T)$). As the lattice distortion is zero in the tetragonal phase, the onset of superconductivity is not captured in $|m_{x} - m_{y}|(T)$. In our model calculations, we observe a field-induced low-temperature increase of the orthorhombic distortion, like that observed in the thermal expansion data for $H\parallel [100]$ (see Fig. \ref{fig:ThermalExpansion}. In the model, this can be attributed to the field-induced weakening of the competition between superconductivity and stripe-type magnetism.

\section{Discussion and Conclusion}

The transition temperatures inferred from our thermal-expansion and heat capacity measurements are summarized in the ($H,T$) phase diagrams of Figs 4d) and 4e) for $H\perp$[001] and  $H\parallel$[001], respectively, and are compared to the theoretical phase diagram shown in Fig. 4(f). Here, $T_2(H)$ is determined experimentally as the position of the maximum in heat capacity (light blue arrow in Fig. \ref{fig:spec_heat}) and in the coefficient of linear thermal expansion $\alpha_{110}(T,H)=1/L_{100}((\partial\Delta L_{110})/\partial T)$. As the peak observed in specific heat broadens considerably with increasing field in the [001] direction, this criterion is restricted to fields less than 6 T. On the other hand, we determine $T_c(H)$ as the maximum of the mean-field discontinuity in heat capacity, rather than using an entropy-conserving construction, because it is partially obstructed by the nearby broadened transition to the low temperature orthorhombic single-Q state. The small offset in $T_2(H)$ between the heat capacity (light blue circles) and thermal-expansion data (light blue squares) is likely related to slight variations in $\mathrm{K}$ content between the two samples, but this does not affect our discussion since both lines are found to run parallel till 10 T.

We find that $T_c$ is weakly suppressed for both field orientations with $((\partial T_c)/\partial H)\approx 0.1$K/T for $H\perp$[001] with the superconducting anisotropy being $\approx 2$, a typical value also for other Fe-based superconductors \cite{Tanatar2017, Willa2019, Smylie2019a, Willa2020, Hardy2020}. In contrast, the suppression of $T_2 (H)$ is more pronounced, about a factor 4 larger than that of $T_c (H)$, revealing that a minor weakening of $T_c (H)$ has a strong impact on the transition into the reentrant single-Q phase. Furthermore, the anisotropy of suppression of $T_2 (H)$ is also close to 2 scaling with the superconducting anisotropy which strongly suggests that the stripe-phase reentrance is intimately coupled to superconductivity and by weakening the superconducting order through applied magnetic fields the reentrance of the low temperature single-$Q$ SDW is retarded. Although our model does not account for both crystal and magnetic anisotropies, an anisotropic field dependence of $\alpha_S$ in Eq. (2) matching the experimental superconducting anisotropy immediately reproduces the observed anisotropy of $T_2 (H)$.

Not reproduced by our minimal model is the contrasting thermal expansion behavior of the $b$-orthorhombic axis (see Fig.\ref{fig:ThermalExpansion}) with respect to field direction observed deep inside the reentrant orthorhombic single-Q phase. On the one hand, the model reproduces quite well (in Fig.\ref{fig:simulations}b)) the slight increase of orthorhombicity found for $H\parallel$[110]$_T$. This effect can be understood to result from the well documented competition between superconductivity and the orthorhombicity in the stripe magnetic phase \cite{Nandi2010} i.e. the magnetic field suppresses the superconducting order and thereby promotes the coexisting magnetic order. 
On the other hand, the remarkable rapid disappearance of the orthorhombic distortion and the vanishing entropy discontinuity for $H\parallel$[001], which constitute one of the novel results of our study, are quite puzzling and remain to be understood theoretically. A missing ingredient in our model is the accounting for spin-orbit effects, which are clearly of importance in these materials \cite{Liu2022, Wasser2015, Christensen2015}. For example, it was found that the $c$-axis corresponds to the direction of the magnetic moment in the similar $C_4$-magnetic phase of $\mathrm{Ba}_{1-x}\mathrm{Na}_{x}\mathrm{Fe}_{2}\mathrm{As}_{2}$ \cite{Wasser2015}.
Neutron scattering experiments \cite{Liu2022} have recently shown that spin fluctuations in $\mathrm{Ca}\mathrm{K}(\mathrm{Fe}_{1-x}\mathrm{Ni_x})_4\mathrm{As}_{4}$ are strongly anisotropic, with the $c$ axis contributions dominating over in-plane ones. This finding has in turn been attributed to a strong spin-orbit effect in this material which could be shown to be responsible for the spin-reorientation in the tetragonal double-Q state \cite{Christensen2015}. Anisotropic spin fluctuations might be responsible for the reduced orthorhombicity in the low-temperature $C_2$ magnetic phase for out-of-plane magnetic fields and would have to be considered explicitly in any future model that wants to capture this effect.

In conclusion, we have studied in detail the magnetic field dependence of the transition from the tetragonal double-Q magnetic phase into the orthorhombic, superconducting single-Q phase in {\BaKx} ($x\approx0.25$). We report that a magnetic field suppresses the reentrance of the single-Q orthorhombic phase stronger than superconductivity leading to a splitting of the zero-field first-order transition. The transition into the orthorhombic reentrant phase hereby remains first-order. The suppression rate of this transition temperature is stronger for out-of-plane than for in-plane fields showing the same anisotropy as the superconducting state. This can be reproduced by a phenomenological Ginzburg-Landau model which suggests, that the weakening of superconductivity with magnetic field primarily results in the suppression of the reentrant orthorhombic single-Q phase. Furthermore, we observe a strong reduction of the orthorhombic distortion for out-of-plane fields, that can not be captured with our model and calls for further theoretical investigations.

\begin{acknowledgments}
We acknowledge stimulating discussions with R. Fernandes.
K.\ W. acknowledges support from the Swiss National Science Foundation through the Postdoc.Mobility program.
\end{acknowledgments}

\bibliographystyle{apsrev4-2}

\pagebreak
\appendix

\section{Theoretical Model}\label{app:model}
The phenomenology in {\BaKx} is well captured by the Ginzburg-Landau free energy density $\mathcal{F} = \mathcal{F}_{m} + \mathcal{F}_{s} + \mathcal{F}_{\mathrm{int}}$, as specified in Eqs.~\eqref{eq:Fm}-\eqref{eq:Fint}.
The magnetic texture realizes either a double-$Q$ spin-density wave order ($|m_{x}| = |m_{y}|$) that preserves the $C_{4}$ symmetry or a single-$Q$ stripe magnetic order ($m_{x} m_{y} = 0$) lowering the in-plane rotational symmetry to $C_{2}$. These two possibilities define distinct energy manifolds, between which first order transitions can occur. Once superconductivity appears, the leading interaction term in $\mathcal{F}_{\mathrm{int}}$ only mildly affects the energetics between the two magnetic sectors. The next order term $\propto c_{2}$ on the other hand strongly differentiates between the $C_{4}$ order (where it vanishes) and the $C_{2}$ phase. 

In order to study this model system we limit ourselves to a simple description, where the temperature-dependence is limited to the two mass terms $\alpha$ and $\alpha_{s}$. By identifying $v = \nu x$, with $\nu \approx 32$, we attribute the dominant doping-dependence $x$ to the term dictating the relative energetics of the magnetic phases. Experimentally, the magnetic field-dependence of {\BaKx} can to a large extend be reduced to the superconducting sector. This motivates our choice of coupling only the phenomenological parameter $\alpha_{s}$ to the magnetic field $H$. The specific choice of parameters is provided in Table \ref{tab:params}.
\begin{table}[tbh]
\centering
\noindent\begin{tabular}{|c|C{1.5em}|C{1.5em}|C{1.5em}|C{1.5em}||c|C{1.5em}|c||C{1.5em}|C{1.5em}|}
\hline
  $\alpha(\tau)$	&
  $u$				&
  $g$				&
  $v$				&
  $\gamma$			&
  $\alpha_{s}$		&
  $u_{s}$			&
  $\tau_{c}(H)$		&
  $c_{1}$			&
  $c_{2}$			\\\hline
  $\tau - 1$		&
  4					&
  0.5				&
  $x$				&
  4					&
  $\tau - \tau_{c}$	&
  1					&
  $0.6 - H$		&
  1					&
  0.2 \\\hline
\end{tabular}
\caption{Parameters used in the phenomenological model.
}
\label{tab:params}
\end{table}

Within our model, the calorimetric data can be deduced from $C = T \partial^{2} E / \partial T^{2}$, where $E(T)$ is the system's energy at a given temperature $T$. Furthermore, the coupling of the stripe spin-density state ($|m_{x} - m_{y}| \neq 0$) to the elastic degrees of freedom will result in a length-change of the specimen. We anticipate that $\delta L/L \propto |m_{x} - m_{y}|$ and use the latter as a proxy for the former. 

Note that the orientation of the moments in the different magnetic phases is susceptible to change due to crystal field effects. This aspect is not accounted for within the proposed model. In particular is is not capable of resolving the field-direction dependence at low temperature, see Fig.~\ref{fig:ThermalExpansion}.

\end{document}